\documentclass[a4paper]{article}

\input glyphtounicode.tex %

\usepackage{amsmath}
\usepackage{a4wide}

\ifx\pdftexversion\undefined
\usepackage[dvips]{graphicx}
\else
\usepackage{graphicx}
\usepackage[colorlinks,bookmarksopen,bookmarksnumbered,citecolor=red,urlcolor=red]{hyperref}
\usepackage{xspace} 
\usepackage{tikz}
\usetikzlibrary{shapes,backgrounds,arrows}

\graphicspath{{./}{./Figures/}}

\newcommand{\db}{{\boldsymbol d}\xspace}

\newcommand{\nb}{{\boldsymbol n}\xspace}

\newcommand{\ub}{{\boldsymbol u}\xspace}

\newcommand{\xb}{{\boldsymbol x}\xspace}

\newcommand{\Deltab}{{\boldsymbol\Delta}\xspace}

\newcommand{\nablab}{{\boldsymbol\nabla}\xspace}

\newcommand{\zerob}{{\boldsymbol 0}\xspace}

\newcommand{\determinant}{\textrm{det}\xspace}
\renewcommand{\Dot}{\boldsymbol \cdot}

\newcommand{\Identity}{\textrm{I}}

\newcommand{\transpose}{\mathsf{T}}

\newcommand{\RoundBracket}[1]{\left( #1 \right) }

\newcommand{\BraceBracketLeft}[1]{\left\{ #1 \right. }

\newcommand{\VerticalBracketRight}[1]{\left. #1 \right| }

\newcommand{\fluid}{\mathrm{F}\xspace}
\newcommand{\solid}{\mathrm{S}\xspace}
\newcommand{\interface}{\mathrm{I}\xspace}

\newcommand{\ALE}{{\mathcal{M}}\xspace}

\newcommand{\FSIsolidDeformationGradient}{{\mathrm{G}}\xspace}

\begin{document}

\title{Simulation of left ventricle fluid dynamics with mitral regurgitation from magnetic resonance images with fictitious elastic structure regularization}

\author{T. Lassila$^{\rm a}$ $^{\ast}$, A.C.I. Malossi$^{\rm b}$, M. Stevanella$^{\rm c}$, \\E. Votta$^{\rm c}$, A. Redaelli$^{\rm c}$ and S. Deparis$^{\rm a}$}
\date{Apr 20th, 2014\\
$^{a}${\em{Institute of Mathematics, EPFL, Lausanne, Switzerland}};\\
$^{b}${\em{IBM Research, Z\"urich, Switzerland}}; \\
$^{c}${\em{Bioengineering Department, Politecnico di Milano, Milan, Italy}}}

\maketitle

\begin{abstract}
Computer modeling can provide quantitative insight into cardiac fluid dynamics phenomena that are not evident from standard imaging tools. We propose a new approach to modeling left ventricle fluid dynamics based on an image-driven model-based description of ventricular motion. In this approach, the end-diastolic geometry and time-dependent deformation of the left ventricle cavity are obtained from cardiac magnetic resonance images and a fictitious elastic structure is used to impose the contractile behavior of the left ventricle. This allows seamless treatment of the isovolumic phases. Besides the ventricular motion, the intracavitary fluid dynamics is controlled by the mitral valve. Three different mitral valve models are included in the simulation: an idealized diode (with or without regurgitation) and a lumped parameter model accounting for the opening dynamics of the valve and including regurgitation.

\medskip
{\em Keywords :}
	fluid-structure interaction;  left ventricular fluid dynamics; mitral regurgitation
\end{abstract}

\renewcommand{\thefootnote}{\alph{footnote}}

\section{Introduction}

The high prevalence and rate of mortality of cardiac diseases have driven the development of methods to gain 
quantitative insight into cardiac funtion and, in particular, the function of the left ventricle (LV). Among the 
different aspects of LV biomechanics, intracavitary fluid dynamics plays a pivotal role and can provide a noninvasive 
indicator of different pathological conditions. In healthy LVs, the haemodynamics are characterized by: (i) 
\emph{high intraventricular pressure gradients} during the early diastole (E-wave) that give rise to a 3-D vortex 
ring forming immediately downstream of the mitral valve (MV) leaflets and help generate a strong jet that extends all 
the way to the apex with minimal viscous dissipation; (ii) \emph{sustained rotational flow} during late-diastolic atrial 
contraction (A-wave); and (iii) \emph{rapid contraction} during systole that redirects the blood to the LV outflow tract and 
through the aortic valve. 
These features contribute to optimizing the LV pumping efficiency, as well as to the haemodynamics in the proximal aorta, 
and their alteration is usually associated with pathological conditions.


Computational fluid dynamics simulations can be useful for analyzing LV fluid dynamics to overcome the limitations of current imaging
techniques. Due to the importance of the motion of the LV cavity and the MV leaflets, these fluid dynamics simulations should ideally
be extended to fluid-structure interaction (FSI) simulations taking also into account the interplay between the biological
tissue and moving blood. The setting of a fully realistic FSI model of the LV is particularly challenging, since it has to account for:
i) the complex LV geometry; 
ii) the presence of heart valves, whose modeling is non-trivial and is required to have realistic inflow and outflow boundary conditions;
iii) the complex motion of the LV wall during the cardiac cycle. Such motion can be imposed either directly by means of 
kinematic boundary conditions on the boundary of the LV cavity, or through the explicit modeling of 
the LV myocardium, which requires the modeling of passive and active mechanical properties of myocardial tissue, of the 
laminar structure of the LV and its myocardial fiber architecture, and ultimately of the propagation of contraction. The latter 
approach is more demanding since it requires the identification of a large number of model parameters through complex experimental 
set-ups and procedures, as was done in \cite{Sermesant2011} for patient-specific electromechanical models and in \cite{Wenk2010} for 
valve models. 

Current medical imaging technology can yield the information required to reconstruct LV 3D geometry. For instance, cardiac magnetic resonance 
imaging (cMRI) can be performed with different acquisition sequences to quantify i) LV anatomy, time-dependent volume, 
and wall motion from cine images, ii) regional 2D wall motion and strains from tagged images, iii) local tissue necrosis/fibrosis 
from late gadolinium enhanced images, iv) myocardial fiber architecture from diffusion tensor cMRI, and v) blood velocity 
fields from phase contrast images. These data can be used to feed image-based, patient-specific FSI models, which can be 
used to study different aspects of LV biomechanics by means of morphologically realistic 3D models, as well as to test the 
suitability of different medical and surgical treatments on a patient-specific basis to support clinical planning. Still, 
only cine-cMRI and late gadolinium enhanced acquisitions are routinely performed in clinics, while the acquisition of the 
other sequences is usually limited to research activities due to their complex implementation and their excessive time for 
analysis. 

With increased availability of image-based detailed information, many FSI studies have been performed in realistic LV 
geometries (see \cite{cheng2005fluid,doenst2009fluid,Long2008,nakamura2006influence,Tang2010}), accounting also for LV 
wall electro-mechanics in some cases (see \cite{Nordsletten2011,watanabe2004multiphysics}). In such studies it is
typically assumed that the effect of the MV leaflets is negligible and the MV is modelled by inflow boundary conditions 
that try to represent the time-varying shape and orientation of the MV orifice. In the literature, only recently have appeared
fully three-dimensional fluid dynamics studies in patient-specific LV geometries incorporating the leaflet dynamics
(see \cite{mangual2013comparative,mihalef2011patient,votta2013toward}), 
although a fully realistic model coupling unsteady haemodynamics effects with anisotropic material models for the leaflets 
with contact modelling and inclusion of the effects of chordae tendinae and the papillary muscles still seems out of reach. 
This aspect may appear of minor relevance, but is not: experimental evidence strongly suggests that motion of the MV 
leaflets and LV vorticity influence each other during diastole (see \cite{charonko2013vortices, kim1995}), while computational
studies indicate that the systolic configuration of the anterior MV leaflet plays a role in LV ejection efficiency \cite{dimasi2012influence}.

In this work we present a novel approach to LV FSI modeling. Based on standard short-axis cine cMRI images an in vivo LV 
model-based geometry is reconstructed and regional LV wall 3D displacements are identified from the cMRI by an algorithm 
of \cite{Conti2011}. Space- and time-dependent 3D-displacements are imposed to the LV endocardial surface 
through a thin fictitious elastic solid, so as to avoid difficulties related to exact volume conservation during the isovolumic phases. 
A MV model is introduced through three different approaches. In the first
case the mitral valve is treated as an idealized diode that is either fully open or fully closed and provides no resistance
to the flow. In the second case the ideal diode is modified to allow for regurgitation according to suitable and the valve
resistance. In the third case a lumped parameter model accounts for the opening dynamics of the valve as proposed by 
\cite{Mynard2011} so that the resistance offered by the valve changes in time according to the pressure gradient across it. 
All three models also account for the up-and-down motion of the mitral annulus, but not for its resizing nor the effect of the
immersed leaflets. A comparison between the fluid dynamics predictions is made in order to understand which implications
the choice of the model has on the fluid dynamics predictions.

\section{Methods} \label{sec:methods}

\subsection{Image acquisition and motion reconstruction of the LV}
The methodology was tested on imaging data from a 65-year-old female patient who had a hibernating myocardium in the LAD 
territory and volume overload due to mitral regurgitation; the septal side of the LV was severely akinetic. Using a 
$1.5$~T whole-body Siemens Avantoa MRI scanner, equipped with a commercial cardiac coil, electrocardiogram-gated breath-hold 
cine images of the LV were acquired in multiple short axes using steady state free procession sequences ($20$~time frames/cardiac 
cycle, reconstruction matrix $256\times256$~pixels, in plane resolution $1.719\times1.719$~mm$^2$, slice thickness $8$~mm, 
gap $1.6$~mm). The valves and LV apex were not captured, due to the inherent limitations of the short-axis sequence.
The imaging and reconstruction method are described in detail in \cite{Conti2011}, where a validation of the methodology
was done by comparison to commercial software for CMR analysis, and is briefly summarized here. For every time frame and on 
each short-axis slice the LV endocardial contour was 
semi-automatically detected through the Chan-Vese approach and integrated with a priori knowledge of the statistical 
distribution of gray levels in medical images. Contours were regularized using a curvature-based motion algorithm designed to 
disallow curvatures above the mean Euclidean value. For each time frame, the smooth LV endocardial surface was obtained by 
biplanar cubic spline approximation of previously detected contours. The surface was discretized into approximately $2$\,$000$ 
three-node triangular elements. In the end-diastolic frame, the endocardial surface was divided into six longitudinal sections 
and three circumferential sections, thus obtaining $18$ sectors. For each sector, nine points forming a $3\times3$ mapped grid 
were identified and the corresponding local principal curvatures calculated as in \cite{Vieira2005}. Then, each point of each 
section was tracked throughout the subsequent time-points by means of a nearest neighbor search, based on the minimization of 
the frame-by-frame variations in spatial position and local curvature. The result of this procedure was the time-dependent position of 
$84$ landmark points for the entire endocardial surface.

\subsection{Patient-specific LV geometry for FSI simulations}

A model-based approach for constructing a computational LV geometry was used. The LV intracavitary volume mesh was obtained 
by starting from an idealized tetrahedral mesh representing the general LV shape and then morphing it by nonrigid deformations 
to fit the short-axis landmark points in the end-diastolic configuration. To mimic the LV we used a truncated ellipsoid with short extruded 
sections extending from both valves, which were modelled as ellipsoidal surfaces. By an extrusion procedure a thin fictitious 
elastic structure around the endocardium was generated for the purposes of imposing the motion of the LV. The dimensions and
alignment of the mitral valve long and short axis were fitted to those observed in the long-axis sequence. The mitral valve
annulus was approximated by an ellipsoid with major axis $3.5$~cm and minor axis $2.6$~cm at peak systole, whereas the aortic 
valve was approximated by a circle of diameter $1.8$~cm.

The idealized LV was aligned and resized to match the position of the landmarks at end-diastolic configuration of the 
anatomically correct LV reconstructed from MRI. First, we applied a rigid transformation to the idealized LV to align 
the vertices on the endocardium surface with the MRI-derived set of landmark points. Then, we defined a least-squares 
error functional measuring the discrepancy between the two sets of points and solved a least-square minimization 
problem to find the optimal scaling of the idealized LV in each of the three major axis directions.  During this minimization 
process the volume of the LV was constrained to equal the MRI-based approximation and the total length of the LV was
constrained to equal that observed in the long-axis sequence. The top-most short-axis slice was assumed to be located 
at a distance of $2$~cm of length from the valvular plane.

\begin{figure}
	\centering
	\includegraphics[width=\textwidth]{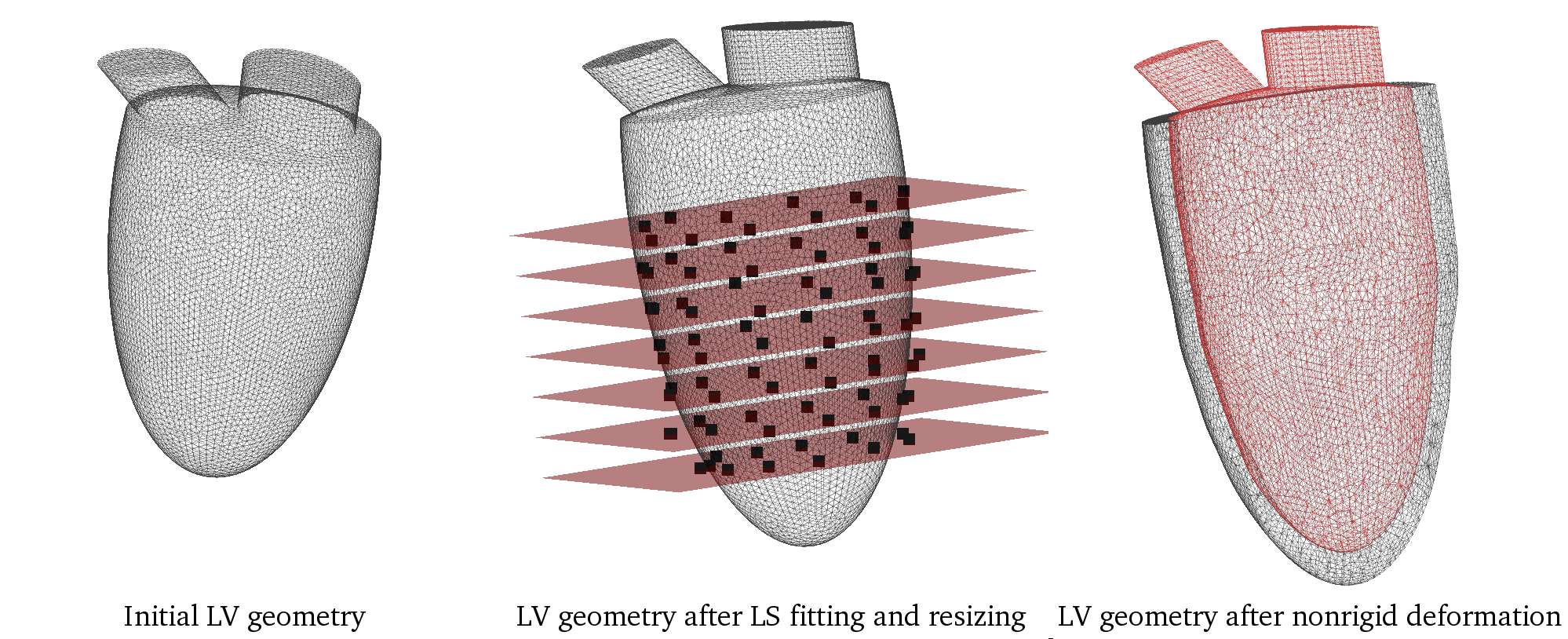}
	\caption{The idealized ellipsoidal LV geometry (left) is first aligned to the short-axis planes and landmarks. An affine 
	transformation that minimizes the discrepancy between the landmarks and the endocardial surface is then sought by least-squares 
	fitting (middle). In the final step a nonrigid radial basis function deformation is applied to
	both fluid and solid geometries to obtain the computational mesh (right).}
	\label{fig:fitting}
\end{figure}

After alignment and resizing, the ideal LV geometry was nonrigidly deformed to fit the landmark points.
First, for each landmark point we identified the closest boundary mesh nodal point on the endocardium at end-diastole.
Using this point-to-point identification an initial deformation was applied to the idealized LV using radial basis 
function interpolation that warped the idealize LV endocardium to match the landmark points. This deformed configuration 
was then taken as the initial end-diastolic configuration. After the nonrigid deformation was applied to obtain the 
end-diastolic configuration that matched the position of the landmarks at the end-diastolic instant, the motion of the 
landmark points was extrapolated and used to drive the finite element simulation of the ventricular haemodynamics. The 
positions of the landmark points on the endocardium were chosen as interpolation centers, and the motion of the 
landmarks in space and time was extrapolated in the space-time domain by performing radial basis function interpolation 
in space and trigonometric interpolation in time. In order to regularize the motion, the five highest temporal modes 
were neglected so that the resulting LV volume reconstruction was monotone increasing during diastole. This resulted in 
a globally smooth and time-periodic extension field of the LV motion throughout one heartbeat. The procedure is illustrated
in Fig.~\ref{fig:fitting}.


A separate long-axis sequence was used to register the mitral leaflets relative to the position of the aortic 
root. Two phases of the long-axis sequence were used for the mitral leaflet registration at diastolic (fully open) 
and systolic (closed but regurgitant) positions. This was performed manually by identifying $42$--$46$ landmark points split 
between the two leaflets and least-squares fitting of two bivariate polynomial surfaces of total degree four
for both leaflet surfaces. From this reconstruction the maximum opening area of the mitral valve was estimated
at $7.22$~cm$^2$ and the mitral regurgitant area was estimated at $0.07$~cm$^2$. 

\subsection{Finite element modelling of LV haemodynamics}

Ventricular haemodynamics were modelled with a finite element FSI model. The fluid part 
of the FSI modeling problem consisted of the Navier-Stokes equations (NSE) for incompressible Newtonian fluids written 
in the arbitrary Lagrangian-Eulerian (ALE) formulation. The thin structure surrounding the LV was only used to impose 
the motion of the LV in such a way as to obtain smooth LV pressure fields both in space and time, and was thus 
modelled as a thin pseudo-incompressible linear elastic and isotropic material. In order to impose the reconstructed
motion of the LV in the haemodynamics simulation, we used the extrapolated space-time motion field as a boundary 
condition on the external surface of the thin extruded structure surrounding the LV. The current study only focuses
on LV haemodynamics and so no predictions of myocardial strains or stresses were needed, though these could have been 
obtained by replacing the thin linear isotropic structure with a physiologically motivated nonlinear orthotropic 
material of realistic thickness. While there is extra computational cost related to the fluid-solid coupled problem, 
this formulation allowed both the recovery of a spatio-temporally smooth LV pressure field as well as the ability 
to seamlessly simulate both isovolumic phases, which are a known difficulty for the pure NSE-in-moving-domains 
-formulation (e.g. \cite{chnafa2014image} eliminate the isovolumic phases completely in an otherwise high-fidelity
simulation of full-heart fluid dynamics).

The resulting FSI problem reads
\begin{equation}
\BraceBracketLeft{\begin{array}{r@{\,\,}c@{\,\,}l@{\quad}l}
\VerticalBracketRight{\dfrac{\partial \ub_\fluid}{\partial t}}_{\xb^0} + \RoundBracket{\RoundBracket{\ub_\fluid - \VerticalBracketRight{\dfrac{\partial \db_\fluid}{\partial t}}_{\xb^0}} \Dot \nablab} \ub_\fluid -\dfrac{1}{\rho_\fluid}\nablab \Dot \sigma_\fluid &=& \zerob & \textrm{in } \Omega^t_\fluid \times (0,T],\\[2ex]
\nablab \Dot \ub_\fluid  &=&  0 & \textrm{in } \Omega^t_\fluid \times (0,T],\\[1ex]
\rho_\solid \dfrac{\partial^2 \db_\solid}{\partial t^2} - \nablab \Dot \sigma_\solid &=& \zerob & \textrm{in } \Omega^0_\solid \times (0,T],\\[2ex]
-\Deltab \db_\fluid &=& \zerob & \textrm{in } \Omega^0_\fluid \times (0,T],\\[1ex]
\ub_\fluid \circ \ALE^t - \dfrac{\partial \db_\solid}{\partial t} &=& \zerob & \textrm{on } \Gamma^0_\interface \times (0,T],\\[2ex]
\sigma_\solid \Dot \nb_\solid - J_\solid \FSIsolidDeformationGradient_\solid^{-\transpose} \RoundBracket{\sigma_\fluid \circ \ALE^t} \Dot \nb_\solid   &=& \zerob & \textrm{on } \Gamma^0_\interface \times (0,T],\\[1ex]
\db_\fluid - \db_\solid &=& \zerob & \textrm{on } \Gamma^0_\interface \times (0,T],
\end{array}}
\label{eq:GlobalFSI}
\end{equation}
where $(0,T]$ is the time interval, $\ub_\fluid$ the fluid velocity, $\rho_\fluid$ and $\rho_\solid$ are 
the fluid and solid density, respectively, $\nb_\solid$ is the outgoing normal direction applied to the 
solid domain, $\FSIsolidDeformationGradient_\solid = \Identity + \nablab \db_\solid$ the solid deformation 
gradient (with $\Identity$ the identity matrix), and $J_\solid = \determinant\RoundBracket{\FSIsolidDeformationGradient_\solid}$. 
In addition, $\sigma_\fluid$ and $\sigma_\solid$ are the Cauchy and the first Piola--Kirchhoff stress 
tensors. The motion of the interior vertices of the fluid mesh was obtained by harmonic extension from 
the FSI interface by solving an elliptic PDE.

\subsection{Modelling of insufficient mitral valve dynamics} 

For the purposes of this study we performed the LV haemodynamics simulations with three different models
for the mitral valve.

\textbf{Model A:} The classical model for cardiac valves is the ideal diode model, which offers no resistance
to the blood flow and opens and closes instantaneously in response to the changing of the pressure gradient
sign and flow direction:
\begin{equation}  \label{eq:diode_valve}
	Q_{\textrm{mv}} = 
	\left\{	
	\begin{aligned}
		\frac{p_{\textrm{pv}} - p_{\textrm{lv}}}{R_{\textrm{la}}}, \quad &\textrm{ if } p_{\textrm{pv}} > p_{\textrm{lv}} \\
		0, \quad &\textrm{ if } p_{\textrm{pv}} \leq p_{\textrm{lv}} \\
	\end{aligned}
	\right. ,
\end{equation}
where $p_{\textrm{pv}}$ and $p_{\textrm{lv}}$ are the pulmonary and LV pressure respectively. In this model the mitral 
inflow rate $Q_{\textrm{mv}}$ was imposed as a boundary condition on the FSI LV problem using Lagrange multipliers
to enforce the defective boundary condition (see e.g.~\cite{Formaggia2002}),
which has the benefit that no explicit velocity profile needs to be imposed at the inflow. In order to stabilize the
velocity at the mitral valve due to flow reversal effects, the tangential component of the velocity field at the inlet 
was further constrained to zero (see \cite{moghadam11} and the discussion therein), leading to the inflow boundary condition:
\begin{equation}
	\int_{\Gamma_{\textrm{in}}} \ub_\fluid \cdot \nb \: d\Gamma = 0, \quad (\textrm{I} - \nb\nb^T) \ub = \boldsymbol{0} \textrm{ on } \Gamma_{\textrm{in}}.
\end{equation}

\textbf{Model B:} Is an extension of Model A that incorporates regurgitation and inertial effects of the valve on the
fluid dynamics. In this model the flow rate is given by the Bernoulli's equation for flow through an orifice:
\begin{equation} \label{eq:bernoulliValve}
	p_{\textrm{pv}} - p_{\textrm{lv}} = R_{\textrm{la}} Q_{\textrm{mv}} + B Q_{\textrm{mv}} |Q_{\textrm{mv}}| + L \frac{dQ_{\textrm{mv}} }{dt}
\end{equation}
where $B = \rho / (2 A_{\textrm{eff}}^2)$ is the Bernoulli resistance of the valve and
$L = \rho \ell_{\textrm{eff}} / A_{\textrm{eff}}$ the blood inertance. The coefficients 
$L$ and $B$ are determined by the effective orifice area $A_{\textrm{eff}}$ that switches
between open and closed valve configurations similarly to the ideal diode case:
\begin{equation}
	A_{\textrm{eff}} = 
	\left\{	
	\begin{aligned}
		A_{\max}, \quad &\textrm{ if } p_{\textrm{pv}} > p_{\textrm{lv}} \\
		A_{\min}, \quad &\textrm{ if } p_{\textrm{pv}} \leq p_{\textrm{lv}} \\
	\end{aligned}
	\right. .
\end{equation}
For $A_{\min} > 0$ the model allows regurgitation to take place. 
The maximum and minimum orifice area were calibrated from the long-axis segmentation of the valve geometry and were estimated as 
$A_{\max}=7.22$~cm$^2$ and $A_{\min}=0.07$~cm$^2$  respectively, for the case studied. A numerically stable time discretization 
was obtained for \eqref{eq:bernoulliValve} by using the semi-implicit scheme 
\begin{equation}
	Q^{n,k}_{\textrm{mv}} = \frac{Q^{n-1}_{\textrm{mv}} + \frac{\Delta t}{L} \left( p_{\textrm{la}}^{n,k-1} - p_{\textrm{lv}}^{n,k-1} \right)}{1 + \Delta t B / L 
	\, |Q^{n-1}_{\textrm{mv}}|}.
\end{equation}
In this model the pressure $p_{\textrm{lv}}$ was imposed as a normal stress boundary condition on the FSI LV problem along
with the aforementioned tangential velocity stabilization condition:
\begin{equation} \label{eq:bcModelB}
	\left(\sigma_\fluid + p_{\textrm{lv}} \textrm{I} \right) \nb = \boldsymbol{0} \textrm{ on } \Gamma_{\textrm{in}}, \quad (\textrm{I} - \nb\nb^T) \ub = \boldsymbol{0} \textrm{ on } \Gamma_{\textrm{in}}.
\end{equation}

\textbf{Model C:}
To model more precisely the valve opening dynamics we used a lumped parameter model proposed by 
\cite{Mynard2011}, which prescribes simple and smooth opening and closing dynamics for $A_{\textrm{eff}}$ 
without explicitly modeling the valve leaflets. In this model the flow rate through the mitral valve is 
again given by Bernoulli's equation \eqref{eq:bernoulliValve} for flow through an orifice, which in turn 
depends on an internal variable $\zeta \in [0,1]$ according to
\begin{equation}
	A_{\textrm{eff}}(t) = \left[ A_{\max} - A_{\min} \right] \zeta(t) + A_{\min},
\end{equation}
where the internal variable evolves according to the rate equation
\begin{equation} \label{eq:leaflet_momentum_equation}
	\dfrac{d\zeta}{dt} = 
	\left\{
	\begin{aligned}	
		(1-\zeta) K_{vo} \left( p_{\textrm{la}} - p_{\textrm{lv}} \right), &\quad \textrm{ if } p_{\textrm{la}} \geq p_{\textrm{lv}} \\
		\zeta K_{vc} \left( p_{\textrm{la}} - p_{\textrm{lv}} \right) 	 , &\quad \textrm{ if }	p_{\textrm{la}} \leq p_{\textrm{lv}}
	\end{aligned}
	\right. .
\end{equation}
This model captures the valve opening dynamics and represents mitral insufficiency, but does not model the 
effect of the leaflets on the local flow pattern. The boundary conditions applied on the FSI LV problem
were identical to \eqref{eq:bcModelB}.
 
\subsection{Ventricular pre- and afterload}

A scenario of chronic mitral regurgitation with a constant pulmonary pressure of $p_{\textrm{pv}} = 10$~mmHg 
was chosen for the preload. 
For the ventricular afterload we used a standard three-element windkessel model with the parameters given by
\cite{Stergiopulos1999} to determine the aortic flow rate $Q_{\textrm{ao}}$ and aortic pressure $p_{\textrm{ao}}$ as:
\begin{equation} \label{eq:windkessel3}
	\frac{dQ_{\textrm{ao}}}{dt} = \frac{1}{R_{\textrm{ao}} R_{\textrm{pe}} C_{\textrm{ao}}} \left[ R_{\textrm{pe}} C_{\textrm{ao}} \frac{d}{dt}(p_{\textrm{ao}} - p_{\textrm{ve}}) + (p_{\textrm{ao}} - p_{\textrm{ve}}) - (R_{\textrm{pe}} + R_{\textrm{ao}}) Q_{\textrm{ao}} \right],
\end{equation}
where $R_{\textrm{ao}}$ is the aortic resistance, $R_{\textrm{pe}}$ is the peripheral resistance, and
$C_{\textrm{ao}}$ is the aortic compliance. The venous pressure $p_{\textrm{ve}}$ was fixed at $5$~mmHg.
All of values of the model parameters used are listed in Table~\ref{Tab:HeartParameters}.
Model A was used for the aortic valve in all three cases.
It is also possible to model the aortic valve similarly to the mitral one, though this was not deemed
necessary in the presence of a healthy aortic valve. For the coupling algorithm between the 3D LV model 
with the windkessel model for the ventricular afterload, we refer to 
\cite{Malossi2011,Malossi2011Algorithms1D,Malossi2011Algorithms3D1DFSI}. 

\section{Results}

The LV FSI simulation was initialized at rest with zero velocity and pressure and driven for a few heartbeats at 75 bpm
until pressure conditions stabilized into periodicity. The pulmonary pressure was ramped to $10$~mmHg in
the course of the first $100$~ms of the simulation to provide an impulse-free initialization, then kept constant
for the rest of the run. No further initializations or regularizations needed to be performed. A fixed time step 
of $1$~ms was used for the solution of the 
FSI problem with second-order backward differentiation formula in time. The finite element problem was discretized 
using 124$\,$942 tetrahedral elements in the fluid domain and piecewise linear basis functions for both velocity 
and pressure approximation. Well-posedness of the problem was guaranteed by convective and pressure stabilization
performed with the interior penalty method by \cite{Burman2006}. The peak Reynolds number inside the LV during the 
diastolic phase was around $2$\,$000$, indicating transitional but not fully turbulent flow, and therefore no turbulence 
modelling was performed.

\begin{figure}
	\centering
	\includegraphics[height=8cm]{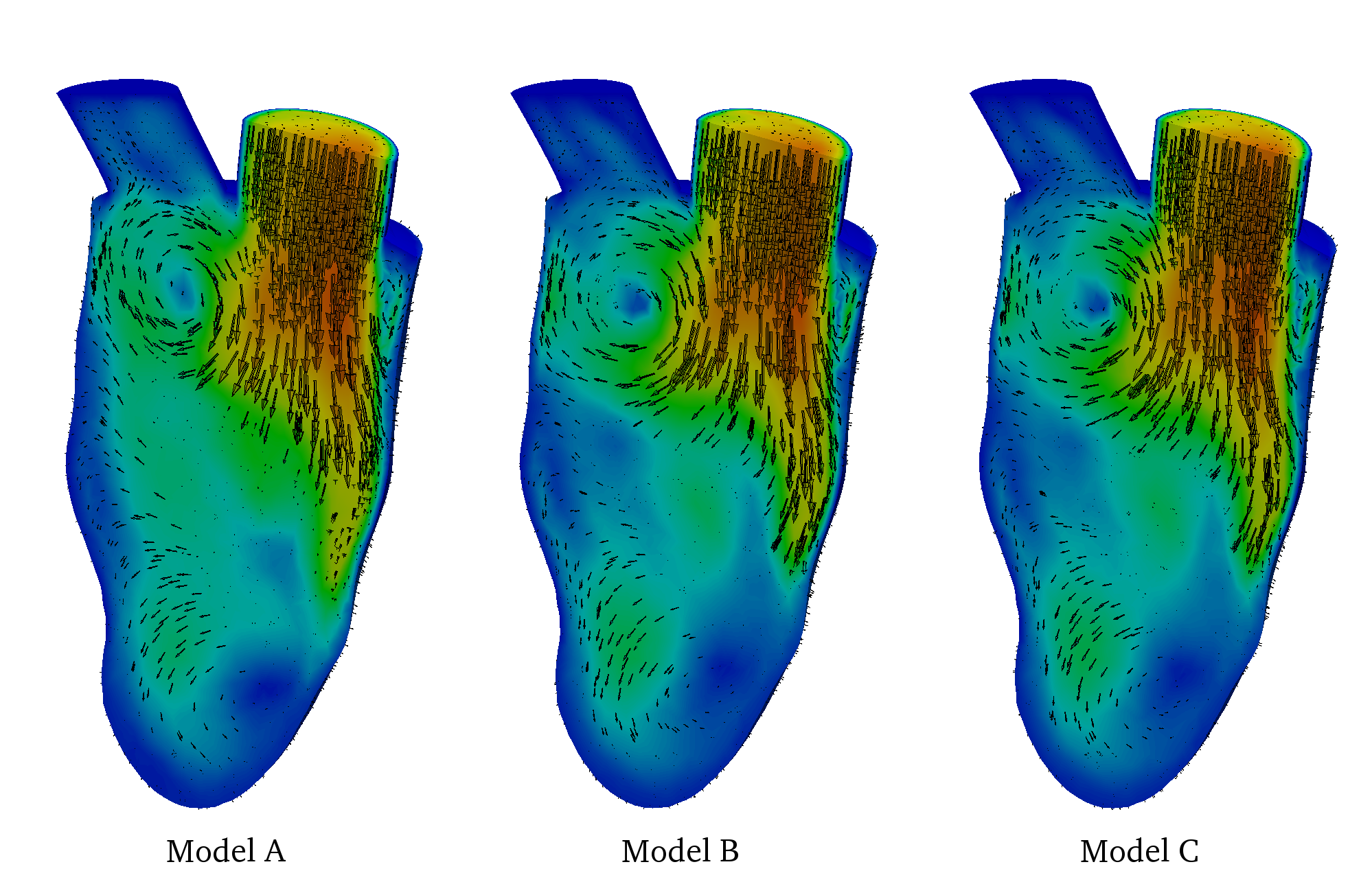}
	\includegraphics[height=8cm]{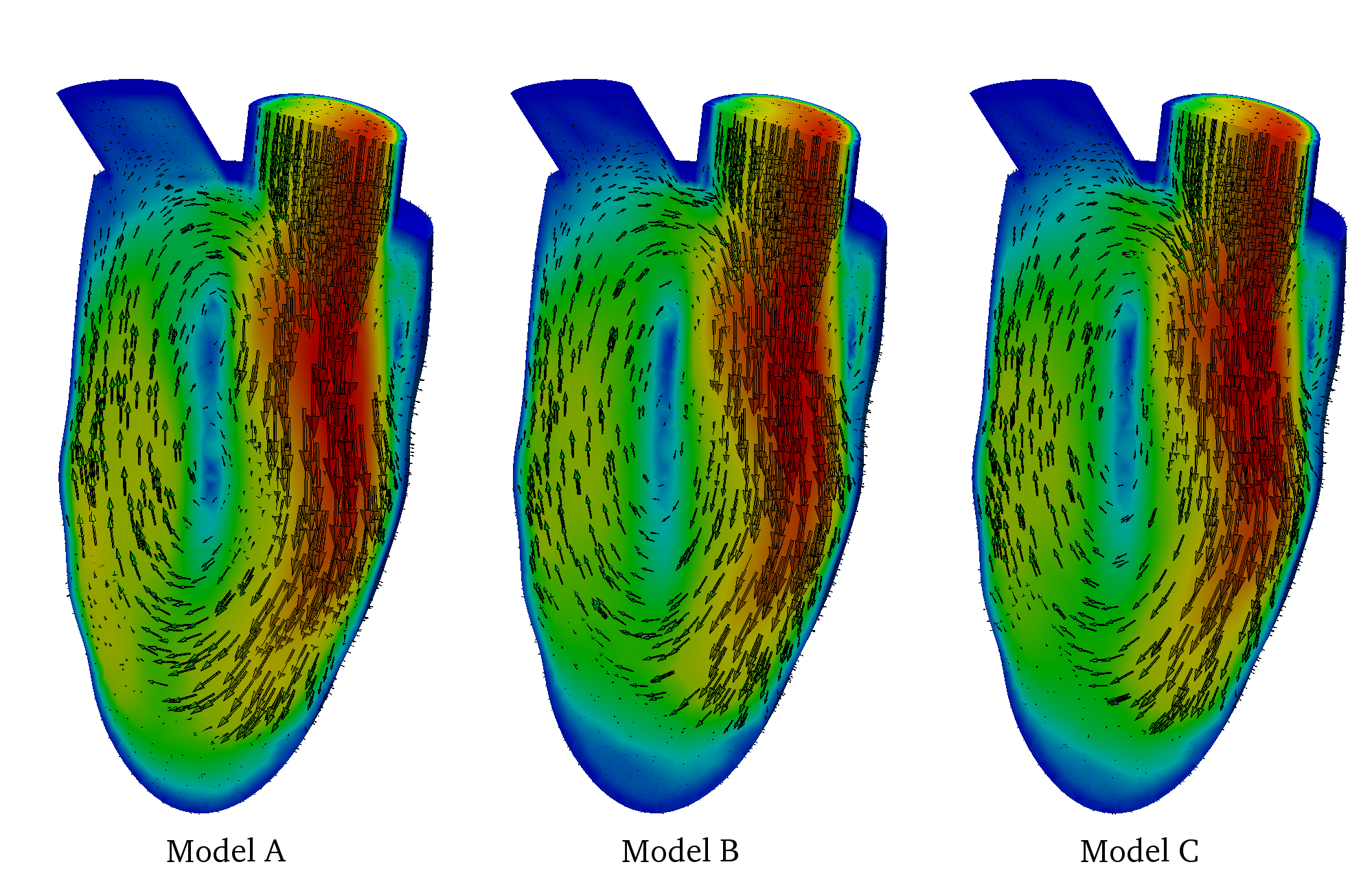}	
	\includegraphics[height=8cm]{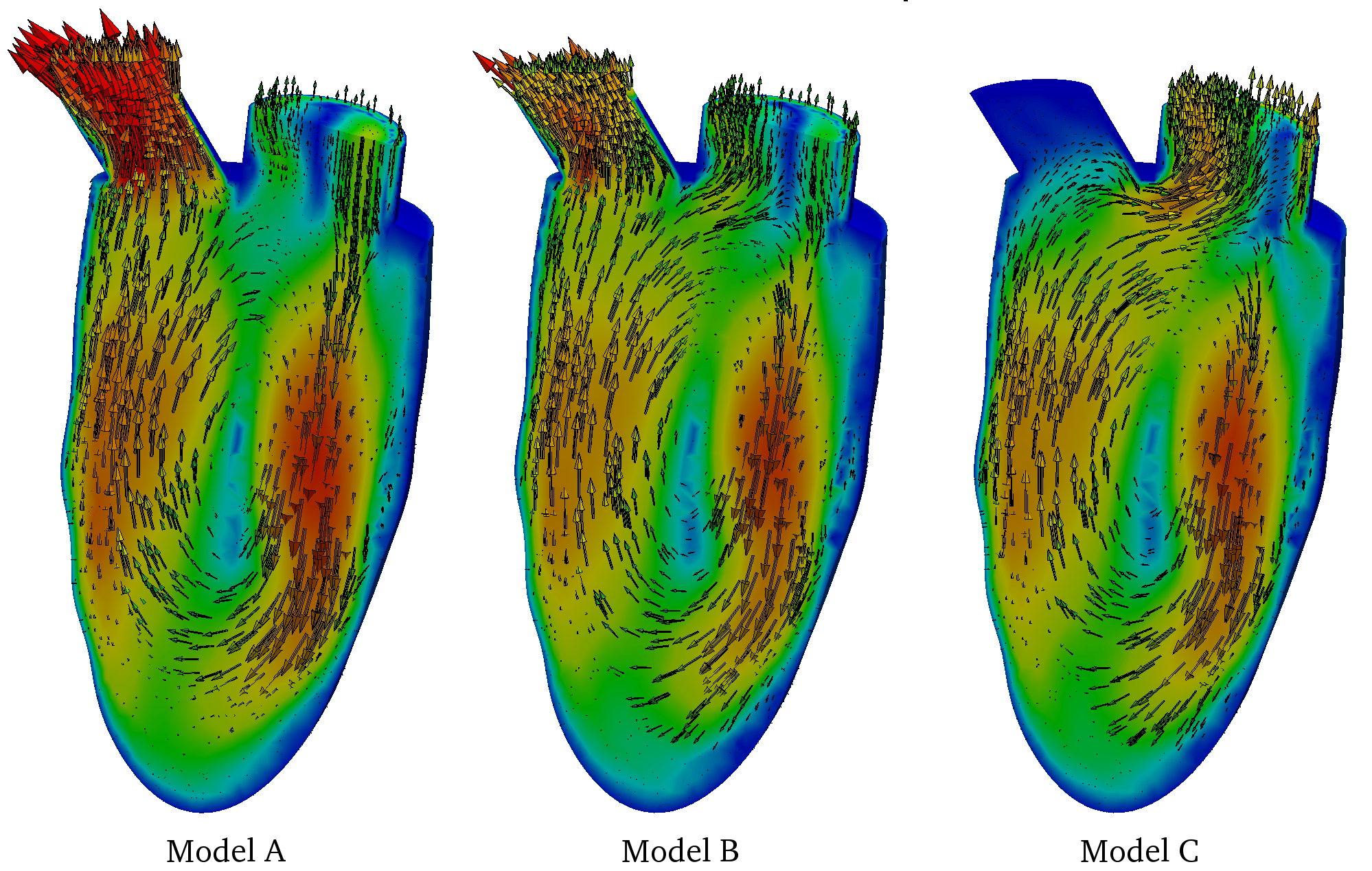}		
	\caption{Comparison of vortex jets for the Models A, B and C. Top row: early diastolic velocity ($t = 550$~ms). 
	Middle row: late diastolic velocity ($t = 750$~ms). Bottom row: early systolic velocity ($t = 800$~ms). Color bar
	ranges between $0-40$~cm/s.}
	\label{fig:vortices}
\end{figure}

The velocity field and the diastolic jet for each of the three different valve models A, B and C is presented
in Fig.~\ref{fig:vortices} for three different time instances: early diastole, late diastole and early
systole. In all three cases the diastolic jet is strongly driven towards the lateral wall and generates a large 
vortex near the aortic root that expands to fill the entire LV during the late diastolic A-wave. These features 
are independent of the inflow boundary condition applied (flow rate in Model A and pressure in Models B and C)
and the opening dynamics of the MV. All three models exhibited vorticial flow at the mitral inlet during early
systole, but the numerical simulation remained stable and convergent throughout, indicating a successful stabilization 
of the inlet boundary condition during flow reversal.

\begin{figure}[h]
	\centering
	\includegraphics[height=4.25cm]{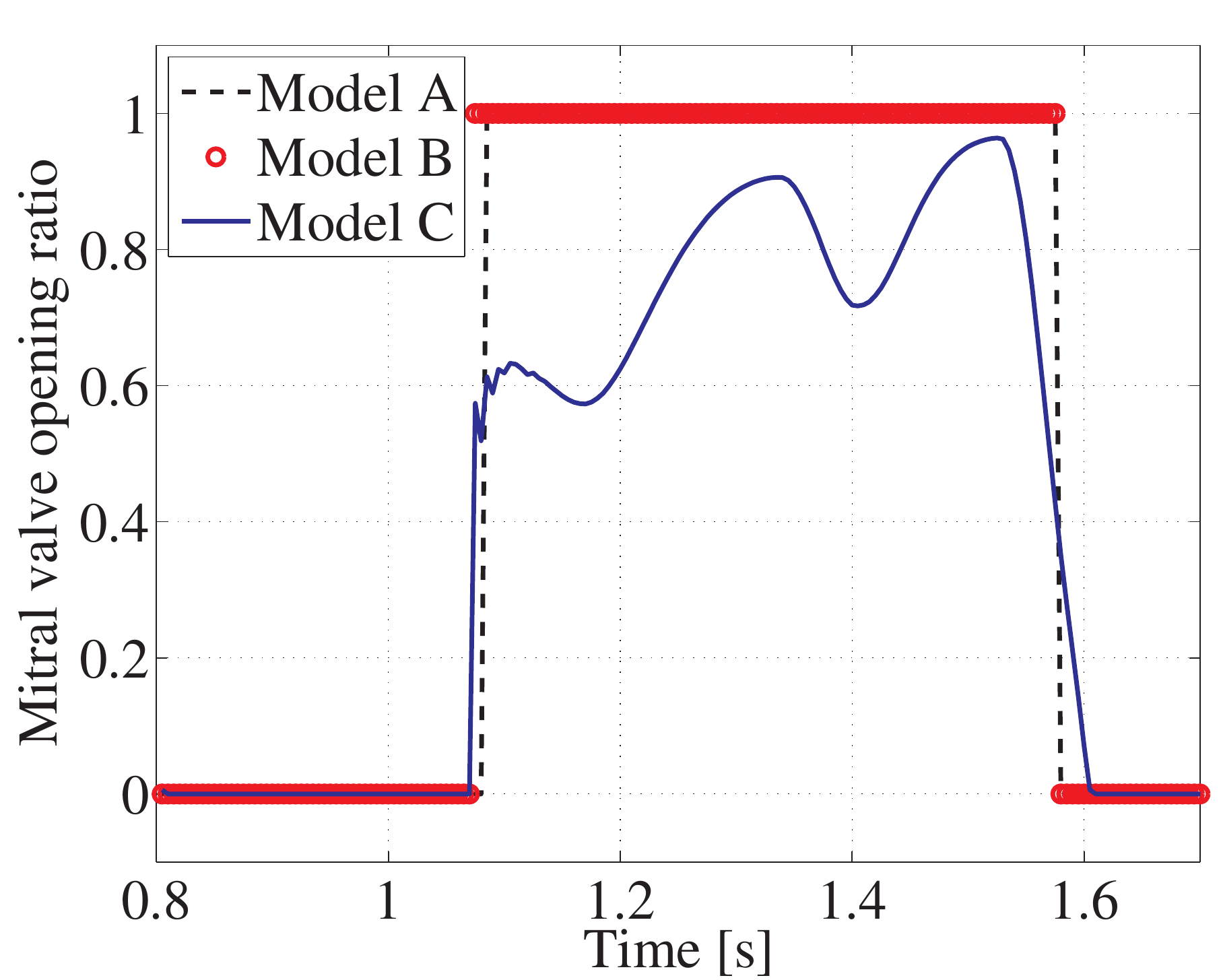}
	\includegraphics[height=4.25cm]{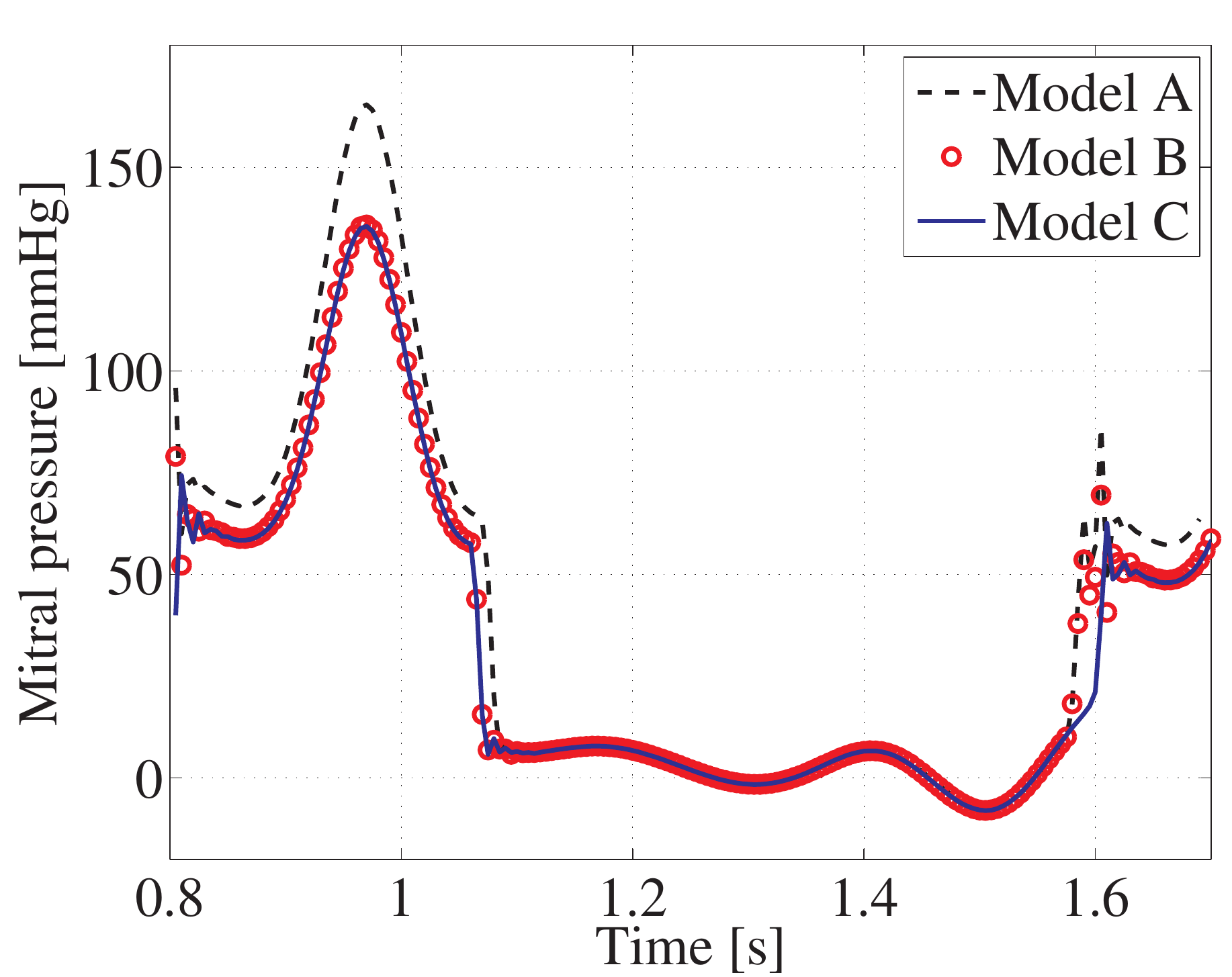}
	\includegraphics[height=4.25cm]{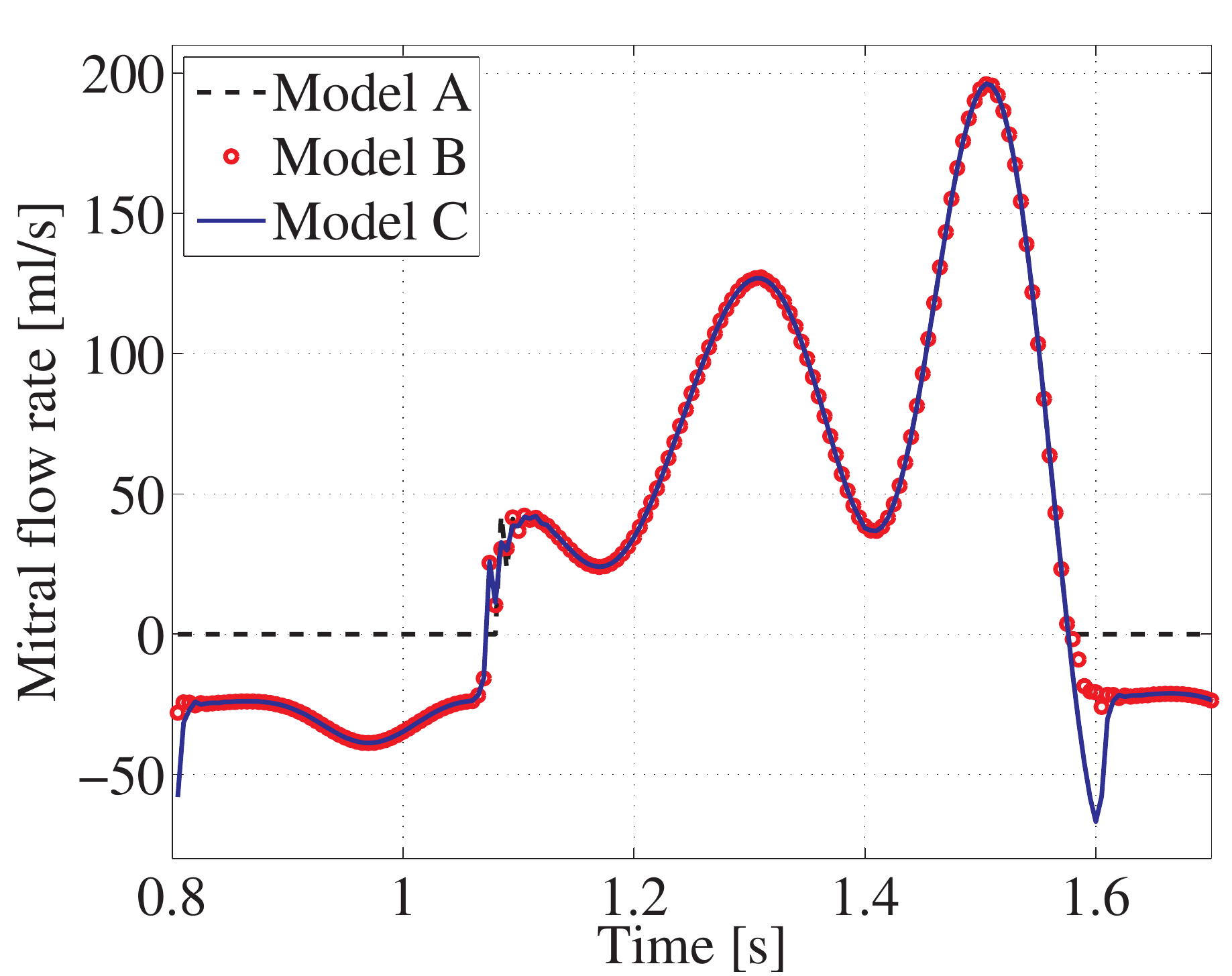}		
	\caption{Comparison of mitral valve characteristics for the valve Models A, B and C. The absence of mitral regurgitation
	in Model A leads to an increase of 19\% in the prediction of systolic pressure peak pressure. The difference between
	the simple regurgitant valve (Model B) and the dynamic regurgitant valve (Model C) is negligible in terms of pressure and flow rate.}
	\label{fig:valve_behavior}
\end{figure}

The mitral valve opening ratio, LV pressure and LV volume in time for the three different valve Models A, B and C 
are presented in Fig.~\ref{fig:valve_behavior}. Model A stands apart from the other two due to the absence of
mitral regurgitation, leading to larger systolic pressure and delayed opening of the MV by about $10$~ms. 
The MV inflow is strongly bimodal and the A-wave is considerably stronger than the E-wave, which is consistent
with clinical findings of chronic or compensated mitral regurgitation when the left atrium has to compensate for the diminished
filling of the LV.
Again, very little quantitative difference between Models B and C can be observed in terms of pressure and flow rate.
This can be explained by the fact that the inflow/outflow volumetric flow rates are largely constrained by the imposed 
motion of fictitious elastic structure that follows from the 4-D reconstruction.

Table~\ref{Tab:MRIndicators} shows the regurgitant volume and its fraction of the total systolic outflow, the
ejection fraction and the viscous dissipation. The predicted regurgitant volume is 9\% higher in Model C, 
mainly due to the slower closure of the mitral valve. Peak viscous dissipation during systole was slightly 
higher in Model A without regurgitation, but almost identical across all three models during diastole. The 
prediction of viscous dissipation depends mainly on whether or not regurgitation is considered or not.

\section{Discussion}
In this work we presented a computational method starting from a standard short-axis MRI sequence 
and proceeding to a 4-D reconstruction of LV motion combined with FSI simulations using
a fictitious elastic structure for regularization. A model-based approach was used to generate the
LV computational geometry. The use of short-axis images permitted a streamlined workflow from
images to simulations with minimal user intervention. The motion of the basal cutplane was added
in order to obtain sufficient ejection fraction and to recover the downward motion of the LV that
was missing from the short-axis images. The mitral annulus diameter orifice area were calibrated
according to manual long-axis segmentation of the LV.

  Compared to standard Navier-Stokes-in-moving-domain formulations, the FSI formulation treats
seamlessly the isovolumic phases without need for explicit volume preservation constraints on the
imposed LV motion. Thus the entire cardiac cycle was simulated in one continuous run without
needing to change boundary conditions or enforcing isovolumic constraints when switching from
systole to diastole and vice versa.

  Comparison of the three different simplified MV models produced qualitatively similar diastolic
flow patterns. A nonsymmetric fluid jet created a two-phase vorticity pattern where a small vortex
was generated near the posterior mitral leaflet in the E-wave and during the A-wave a large
vortex developed to fill the entire LV cavity. Addition of the lumped parameter valve dynamics by
themselves had little effect on the observed LV flow, provided the mitral regurgitation was properly
accounted for, and neither had the change in the LV FSI boundary condition from average flow
rate condition to pressure condition. The fact that consistent results were obtained across all the
valve models used indicates that LV vorticial flow patterns may be simulated to some extent with
just knowledge of the LV wall motion.

  Limitations of the current simulation study include the lack of explicit leaflet modelling and their
effect on the inflow jet, lack of variability of the orifice shape and size, and the missing information
about the orientation of the left atrium with respect to the MV, which influences the diastolic jet
orientation and consequently the vorticity pattern (as shown by \cite{seo2013}). While
the approximate configuration of the mitral valve geometry can be obtained from the MRI in the
fully open and fully closed position, extending this information to the intermediate configurations
and simulating the effect of the leaflet motion to the flow in a proper way requires further work.
Furthermore, solving the problem in the FSI formulation introduces certain additional complexities
in the preconditioning and solution algorithms that may not always be available in commercial
software.

\section*{Acknowledgements}

Clinical patient data for this study was provided by the team of Prof. O.~Parodi at Ospedale Niguarda, Milan, Italy.
S.~Deparis, T.~Lassila, A.~Redaelli, M.~Stevanella, and E.~Votta acknowledge the support of the European Community 
7${}^\textrm{th}$ Framework Programme, Project FP7-224635 VPH2 (`Virtual Pathological Heart of the Virtual Physiological 
Human'). A.~C.~I.~Malossi and S.~Deparis acknowledge the European Research Council Advanced 
Grant `Mathcard, Mathematical Modeling and Simulation of the Cardiovascular System', Project ERC-2008-AdG 227058, as well 
as the Swiss Platform for High-Performance and High-Productivity Computing (HP2C). All of the numerical results 
presented in this paper have been computed using the LGPL \texttt{LifeV} library (\url{www.lifev.org}). All authors 
disclose no conflicts of interest.

{\small
\bibliographystyle{alpha}
\bibliography{references}
}

\begin{table}[h]
\caption{Main parameters of the global problem}
\begin{tabular}{lp{0.45\columnwidth}r@{~}l}
\hline
$\rho_\fluid$      & Blood density                                & 1.060 & g\,cm$^{-3}$ \tabularnewline
$\mu_\fluid$       & Blood dynamic viscosity                      & 0.035 & g\,cm$^{-1}$ s$^{-1}$\tabularnewline
$R_{\textrm{la}}$  & Left atrium resistance                       & 0.09  & mmHg\,s\,cm$^{-3}$ \tabularnewline
$R_{\textrm{ao}}$  & Aortic resistance                            & 0.03  & mmHg\,s\,cm$^{-3}$ \tabularnewline
$R_{\textrm{pe}}$  & Peripheral resistance                        & 0.7   & mmHg\,s\,cm$^{-3}$ \tabularnewline
$C_{\textrm{ao}}$  & Aortic compliance                            & 2     & cm$^3$\,mmHg$^{-1}$ \tabularnewline
$\ell$             & Effective length                             & $1$   & cm \tabularnewline
$K_{\textrm{vo}}$  & Rate coefficient, opening                    & 0.04  & \tabularnewline
$K_{\textrm{vc}}$  & Rate coefficient, closure                    & 0.03  & \tabularnewline
$p_\mathrm{ve}$     & Venous pressure                              & 5     & mmHg\tabularnewline
$p_{\textrm{pv}}$  & Pulmonary pressure                           & 10    & mmHg\tabularnewline
$\rho_\solid$      & Solid density                              & 1.2   & g\,cm$^{-3}$ \tabularnewline
$E_\solid$         & Solid Young's modulus                      & 0.7   & MPa\tabularnewline
$\nu_\solid$       & Solid Poisson's ratio                      & 0.48  &  \tabularnewline
\hline
\end{tabular}
\label{Tab:HeartParameters}
\end{table}

\begin{table}[h]
\caption{Global indicators of mitral regurgitation as predicted by the different models}
\begin{tabular}{c|cccc}
\hline
Model &  Regurgitant & Regurgitant   & Ejection  &  Peak viscous  \tabularnewline
      &  volume [ml] & fraction [\%] & fraction [\%] &  dissipation [mW] \tabularnewline
\hline
A & 0.00 &  0.0\% & 28.9\% & 5.46 (s) / 0.666 (d) \tabularnewline
B & 8.02 & 19.7\% & 29.0\% & 4.21 (s) / 0.657 (d) \tabularnewline
C & 8.73 & 21.5\% & 28.9\% & 4.21 (s) / 0.659 (d) \tabularnewline
\hline
\end{tabular}
\label{Tab:MRIndicators}
\end{table}

\end{document}